\documentclass{moriond}

\def\Journal#1#2#3#4{{#1} {\bf #2}, #3 (#4)}

\def\PRL{\em Phys. Rev. Lett.}
\def\PRD{{\em Phys. Rev.} D}

\def\be{\begin{equation}}
\def\ee{\end{equation}}
\def\bea{\begin{eqnarray}}
\def\eea{\end{eqnarray}}

\newcommand{\Photo}{\includegraphics[height=35mm]{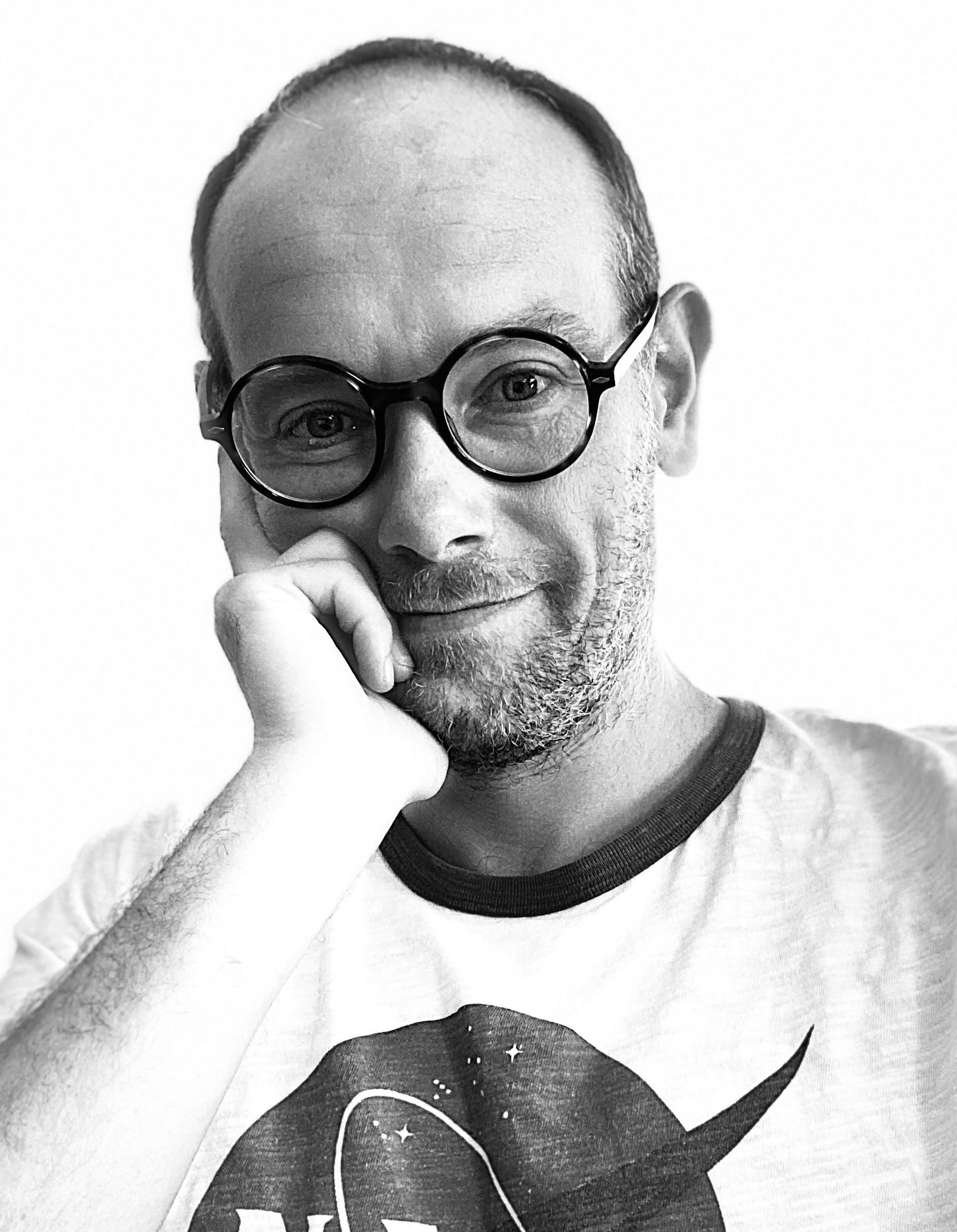}}

\begin{document}
\vspace*{4cm}
\title{On the shortcomings of the Shapiro delay tests of the equivalence principle}

\author{Olivier Minazzoli}

\address{Artemis, Universit\'e C\^ote d'Azur, CNRS, Observatoire C\^ote d'Azur\\ BP4229, 06304, Nice Cedex 4, France}

\maketitle\abstracts{
There are several shortcomings in the ``standard'' Shapiro delay tests of the equivalence principle on cosmological scales \cite{minazzoli:2019pr}. Although many people in the community already acknowledged this in the literature, and proposed alternative ways to compare potential Shapiro delays over cosmological scales---e.g. \cite{bartlett:2021pr,hashimoto:2021pr} and references therein---papers are still submitted to journals with the usual issues.
}

%

\section{``Standard'' definition of the problem}

The ``standard'' Shapiro delay-based test of the equivalence principle relies on the measurement of two arrival times from a unique source but from messengers with different properties, and from the assumption that one can estimate the time of flight of the messengers from indirect observations and a well-suited space-time model. It is meant to test whether or not various messengers propagate along the same trajectories, as it is expected from the Einstein equivalence principle---which encompasses notably both the weak equivalence principle and the Lorentz invariance. In its most common form, it is based on the assumption that the universe is flat at the relevant scale, and that the Shapiro delay $\delta T $ with respect to the propagation time that would have happened in a Minkowski spacetime is given at leading order by, e.g. \cite{krauss:1988pl},
\be
\label{eq:gen_usual_shapiro}
\delta T = - \frac{1+\gamma}{c^3} \int_{\bf r_E}^{\bf r_O} U({\bf{r}}(l)) dl ,
\ee
where $\bf r_E$ and $\bf r_O$ denote emission and observation positions, respectively, $U({\bf{r}})$ is the Newtonian gravitational potential, and the integral is computed along the trajectory. Furthermore, $\gamma$ is usually assumed to parameterize differences for various messengers.

\section{Newtonian potential}

It is often assumed in the literature that the Newtonian potential and its first derivative vanish at infinity, such that it reads
    
    \be \label{eq:standard}
    U=\sum_{P} \frac{G M_{P}}{c^{2}}\left[\frac{1}{\left\|\vec{x}-\vec{x}_{P}\right\|}\right] ,
    \ee
    for a set of point masses with masses $M_P$ and positions $\vec{x}_P$.
    However, in a cosmological context where there are masses everywhere on the past null-cone, this assumption is not physical. As a consequence, it turns out, the Newtonian potential with this definition has a spurious divergence with the number of sources: it quickly no longer can be treated as a perturbation due to the unbounded contribution of all the remote masses.\newline
    
    One can nevertheless show that this divergence can be \textit{renormalized} with an appropriate choice of coordinate time. Among an infinite set of appropriate coordinate times, a convenient one is defined as the proper time of the observer, which is assumed to be at the center of the coordinate system. The potential then instead reads as follows
    
    \be \label{eq:correct}
    U=\sum_{P} \frac{G M_{P}}{c^{2}}\left[\frac{1}{\left\|\vec{x}-\vec{x}_{P}\right\|}-\frac{1}{\left\|\vec{x}_{P}\right\|}-\frac{\vec{x} \cdot \vec{x}_{P}}{\left\|\vec{x}_{P}\right\|^{3}}\right] .
    \ee

This potential does not diverge with the number of sources, and corresponds to the usual potential that is used, for instance, in  cellestial reference frames---such as the ones that are recommended by the \textit{International Astronomical Union}.
\section{Various issues (at several levels)}

\begin{itemize}
        \item \textbf{Eq. (\ref{eq:gen_usual_shapiro}) is not motivated theoretically} $\rightarrow$ There is no reason to expect a modification of the propagation that modifies the Shapiro delay with different $\gamma$ for different messengers.
        \item \textbf{A one-way propagation time is not an observable} $\rightarrow$ It is gauge dependent, as ``\textit{there is no natural way to compare the propagation of a light ray in a curved spacetime with that of a `corresponding' light ray in Minkowski spacetime}'' \cite{gao:2000cq}.
        \item \textbf{Eq. (\ref{eq:gen_usual_shapiro}) neglects cosmology}.
        \item \textbf{Eqs. (\ref{eq:gen_usual_shapiro})+(\ref{eq:standard}) lead to a spurious divergence of the Shapiro delay with the number of sources}.
    \end{itemize}

\section{Outcome of the study Minazzoli et al.}

Even assuming that Eq. (\ref{eq:gen_usual_shapiro}) might make sense for a restricted class of theoretical models that has yet to be found, and for a subset of close enough events, it cannot be used as a fiducial quantity in order to compare the amount of Shapiro delay for various situations---notably because one cannot compute a lower-bound estimate from it \cite{minazzoli:2019pr}. Other approaches that aim to define a fiducial expression might be more appropriate---see, e.g., \cite{nusser:2016ap,bartlett:2021pr,hashimoto:2021pr}.


\section*{References}

\begin{thebibliography}{99}
\bibitem{minazzoli:2019pr} O. Minazzoli et al. \Journal{\PRD}{100}{104047}{2019}
\bibitem{bartlett:2021pr} D.J. Bartlett et al. \Journal{\PRD}{104}{084025}{2021}
\bibitem{hashimoto:2021pr} T. Hashimoto et al. \Journal{\PRD}{104}{124026}{2021}
\bibitem{krauss:1988pl} L.M. Krauss and S. Tremaine \Journal{\PRL}{60}{176}{1988}
\bibitem{gao:2000cq} S. Gao and R.M. Wald \Journal{\em Classical and Quantum Gravity}{17}{24}{2000}
\bibitem{nusser:2016ap} A. Nusser \Journal{\em The astrophysical journal letters}{821}{1}{2016}
%
%
%

\end{thebibliography}
\end{document}